\documentclass{elsart} 

\usepackage[dvips]{graphicx}
\input epsf
\newcommand{\gsim}{\mathrel{\raisebox{-.6ex}{$\stackrel{\textstyle>}{\sim}$} }}

\begin{document}
\begin{frontmatter} 

\rightline{\rm\small IIT-HEP-03/2} 

\title{Muon Cooling Research and Development\thanksref{talk} }
\thanks[talk]{Presented at the {\sl  
International Workshop on Beam Cooling and Related Topics (COOL03)}, May 19--23, 2001, Lake Yamanaka, 
Japan.}
\author{Daniel M Kaplan\thanksref{email}}
\thanks[email]{E-mail: kaplan@fnal.gov} 

\address{Physics Division, Illinois Institute of Technology, 3101 S. Dearborn Street, Chicago, 
Illinois, 60616 USA\\ \vspace{.2in}
for the MuCool Collaboration 
\\ 
} 

\begin{abstract} 

The MuCool R\&D program is described. The aim of MuCool is to develop all key pieces of hardware
required for ionization cooling of a muon beam. This effort will lead to a more detailed
understanding of the construction and operating costs of such hardware, as well as to optimized
designs that can be used to build a Neutrino Factory or Muon Collider. This work is being
undertaken by a broad collaboration including physicists and engineers from many national
laboratories and universities in the U.S. and abroad. The intended schedule of work will lead to
ionization cooling being well enough established that a construction decision for a Neutrino
Factory could be taken before the end of this decade based on a solid technical foundation.  

\end{abstract} 

\end{frontmatter} 





\section{Introduction} 

The possibility of building a muon accelerator has received much attention in recent
years~\cite{Alsharoa,NuFact,FS1,FS2}.  In particular,  decay neutrinos from a stored high-energy
muon beam~\cite{Geer} may well provide the ultimate tool for the study of neutrino
oscillations~\cite{ultimate} and their possible role in baryogenesis via {\em CP}-violating
neutrino mixing~\cite{baryon}. The goal for such a Neutrino Factory is $\sim10^{21}$
neutrinos/year (requiring a similar number of muons/year stored in the ring). 

Muon beams at
the required intensity can only be produced into a large phase space, but affordable existing
acceleration technologies require a small input beam. This mismatch could be alleviated by
developing new, large-aperture, acceleration techniques~\cite{Japan-study}, by ``cooling" the muon
beam to reduce its size, or both. Given the 2.2-$\mu$s muon lifetime, only one cooling technique
is fast enough:  ionization cooling, in which muons repeatedly traverse an energy-absorbing
medium, alternating with accelerating devices, within a strongly focusing magnetic
lattice~\cite{cooling,Neuffer83,Neuffer-COOL03,Kaplan-cooling}. The main focus of the MuCool
Collaboration~\cite{MuCool} has been development of hardware devices and systems that can be used
for ionization cooling of a muon beam.

\section{Ionization cooling channels: design considerations}

In an ionization-cooling channel, ionization of the energy-absorbing medium decreases all three
muon-momentum components without affecting the beam size. This constitutes cooling (reduction of normalized emittance) since the reduction of each particle's momentum results in a
reduced transverse-momentum spread of the beam as a whole. If desired, a portion of this transverse
cooling effect can be rotated into the longitudinal phase plane by inserting suitably shaped energy
absorbers into dispersive regions of the lattice (``emittance exchange"); longitudinal ionization
cooling {\em per se} is impractical due to energy-loss straggling~\cite{Neuffer-COOL03}.

Within an absorber the rate of change of normalized transverse emittance is given approximately
by~\cite{cooling,Neuffer83,Fernow} 
\begin{equation} 
\frac{d\epsilon_n}{ds}\ \approx\ 
-\frac{1}{\beta^2} \left\langle\frac{dE_{\mu}}{ds}\right\rangle\frac{\epsilon_n}{E_{\mu}}\ + 
\ \frac{1}{\beta^3} \frac{\beta_\perp
(0.014)^2}{2E_{\mu}m_{\mu}L_R}\,. 
\label{eq:cool} 
\end{equation}  

Here angle brackets
denote mean value, $\beta$ is the muon velocity in units of $c$, muon energy $E_\mu$ is in GeV, 
$\beta_\perp$ is the lattice beta function evaluated at the location of the absorber, $m_\mu$ is
the muon mass in GeV/$c^2$, and $L_R$ is the radiation length of the absorber medium.

The two terms of Eq.~\ref{eq:cool} reflect the competition between multiple Coulomb scattering of
the muons within the absorber (a heating effect) and ionization energy loss. This favors low-$Z$
absorber materials, the best being hydrogen (see Table~\ref{tab:matl}). Since the heating term is
proportional to $\beta_\perp$, the heating effect of multiple scattering is minimized by
maximizing the focusing strength of the lattice at the location of the absorbers. Superconducting
solenoids are thus the  focusing element of choice in most design studies and can give
$\beta_\perp\sim10$\,cm. The combined effect of the heating and cooling terms sets an equilibrium
emittance at which the cooling rate goes to zero and beyond which a given lattice cannot cool.

\begin{figure} 
\vspace{-.1in} 
\centerline{\scalebox{0.9}{\includegraphics{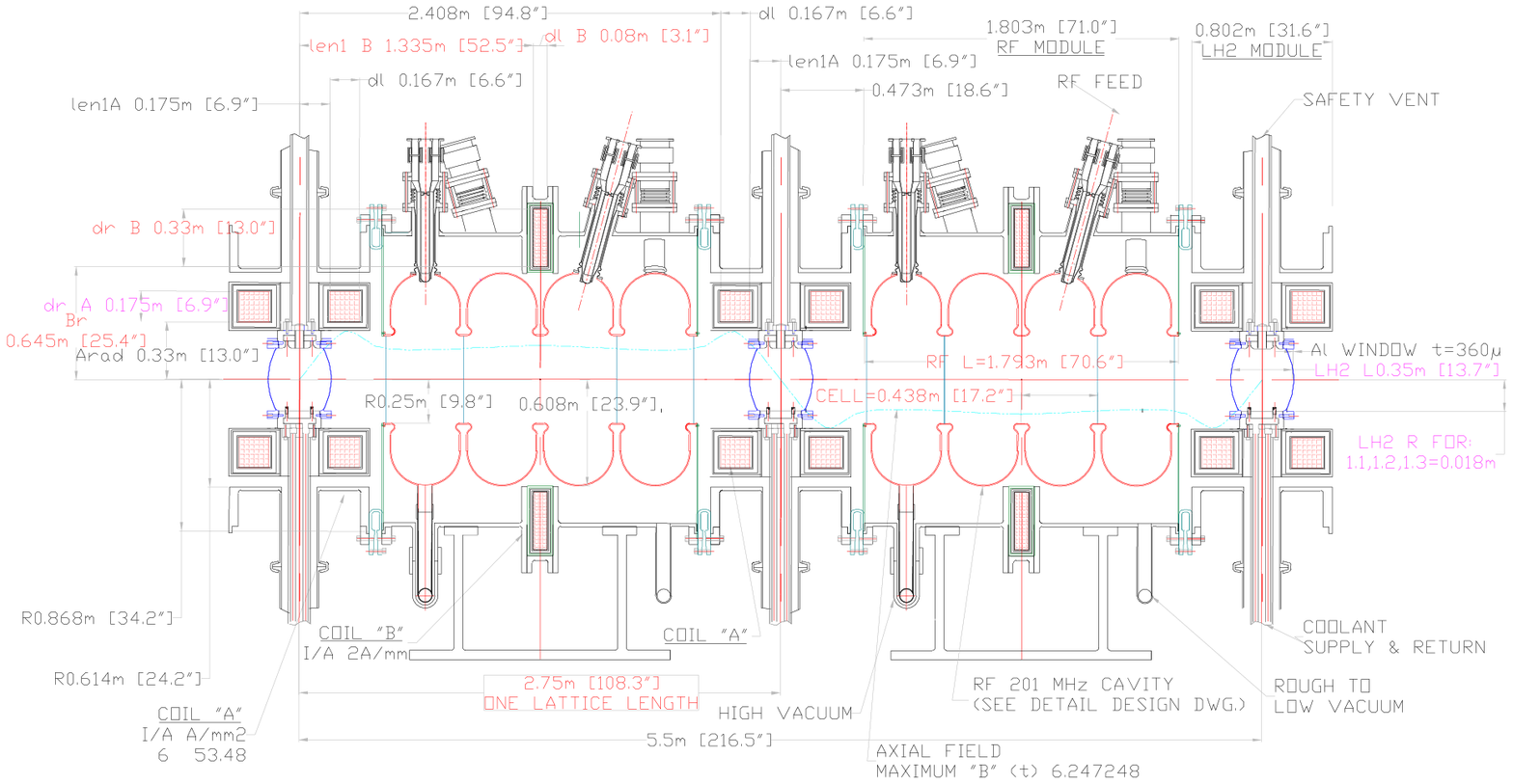}}} 
\vspace{-.4in} 
\caption{Engineering drawing of a section of an ``SFOFO" ionization-cooling lattice (from U.S. 
Neutrino Factory Feasibility Study II~\protect\cite{FS2}). Shown in cross section are three 
liquid-hydrogen absorbers, each enclosed within a pair of ``focusing" solenoids, interspersed 
with two 4-cavity 201-MHz rf assemblies, each encircled by a ``coupling" solenoid.} 
\label{fig:SFOFO} 
\end{figure}

Between absorbers, high-gradient acceleration of the muons must be provided to replace the lost
longitudinal momentum, so that the ionization-cooling process can be repeated many times. Ideally,
this acceleration should exceed the minimum required for momentum replacement, allowing
``off-crest" operation. This gives continual rebunching, so that even a beam with large momentum
spread remains captured in the rf bucket. Even though it is the absorbers that actually cool the
beam, for typical accelerating gradients ($\sim$10\,MeV/m, to be compared with $\langle
dE_\mu/ds\rangle\approx30$\,MeV/m for liquid hydrogen~\cite{PDG}), the rf cavities dominate the
length of the cooling channel (see {\em e.g.}\ Fig.~\ref{fig:SFOFO}). The achievable rf gradient
thus determines how much cooling is practical before an appreciable fraction of the muons have
decayed or drifted out of the bucket.  

 In spite of the relativistic increase of muon
lifetime with energy, ionization cooling favors low beam momentum, both because of the increase of
$dE/ds$ for momenta below the ionization minimum~\cite{PDG} and since less accelerating voltage is
then required.  Most Neutrino Factory and Muon Collider beam-cooling designs and simulations to
date have used momenta in the range $150-400$\,MeV/$c$. (This is also the momentum range in which
the pion-production cross section off of thick targets tends to peak and is thus optimal for muon
production as well as  cooling.) The cooling channel of Fig.~\ref{fig:SFOFO} is optimized for a
mean muon momentum of 200\,MeV$/c$.  

 As a muon beam passes through a transverse ionization-cooling lattice, its longitudinal emittance
tends to grow, due to such effects as energy-loss straggling. The six-dimensional emittance typically
is reduced despite this longitudinal heating. However, if not controlled, the longitudinal heating
leads to beam losses and thus limits the degree of transverse cooling that is practical to achieve.
Cooling lattices with longitudinal--transverse emittance exchange ({\em e.g.}, ring coolers), which
can cool  in all six dimensions simultaneously, have been receiving increasing
attention~\cite{Palmer-ring,Derbenev}. They have the potential to increase Neutrino Factory
performance and decrease cost, and are essential to achieving sufficient cooling for a Muon Collider. 

\section{Muon-cooling technology development} 

An effective ionization-cooling channel must include low-$Z$ absorbers with
(if an intense muon beam is to be cooled) high power-handling capability. To achieve low beta at
the absorbers requires either high solenoidal magnetic field or high field gradient~\cite{quad}.
To pack as much cooling as possible into the shortest distance requires the highest practical
accelerating gradient. The MuCool Collaboration has embarked on R\&D on all three of these
technologies. 

\subsection{High-power liquid-hydrogen absorbers} 

\begin{figure} 
\begin{center} 
\scalebox{.5}{\includegraphics{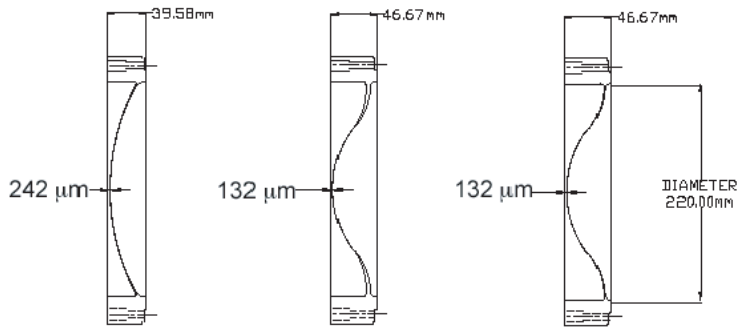}} 
\caption{\label{fig:windows}(Left) first (``tapered torispherical"), (center) second (``bellows"), and
(right) third (``thinned bellows") iterations of custom-shaped and -tapered thin window design.}  
\end{center} 
\end{figure}

The development of high-power liquid-hydrogen (LH$_2$) absorbers with thin windows has been a key goal
of the MuCool R\&D program~\cite{Kaplan-NuFACT01,Kaplan-windows,MACC}.  Simulations as well as theory
show that scattering in absorber windows degrades muon-cooling performance. (This is especially true
for ring coolers, in which the muons circulate until they approach the equilibrium emittance dictated
by multiple scattering.) We have therefore developed new shapes for the ends of the hydrogen flasks
(Fig.~\ref{fig:windows}) that allow significant reduction in their thickness (especially near the
center where the beam intensity is maximum). We have successfully fabricated such tapered, curved
windows out of disks of aluminum alloy using a numerically controlled lathe. We have also devised
novel means~\cite{Kubik} to test these nonstandard windows and demonstrate that they meet their
specifications and satisfy the applicable safety requirements~\cite{MACC}. By optimizing the maximum
stress as a function of pressure, the ``thinned bellows" shape of Fig.~\ref{fig:windows} gives a
central window thickness about one-quarter that of an  ASME-standard ``torispherical"
window~\cite{ASME}. 

Fabrication (and destructive testing as mandated by the Fermilab safety code~\cite{FNAL-safety})
of ``tapered torispherical" windows (Fig.~\ref{fig:windows}) was successfully accomplished previously~\cite{MACC}, using
the 6061-T6 alloy that is standard in cryogenic applications. Fabrication of a series of ``thinned
bellows" windows is currently underway. We are also exploring the use of lithium--aluminum alloys,
such as the 2195 alloy used in the Space Shuttle (80\% stronger than 6061-T6); the resulting
thinness of the window may challenge our fabrication techniques, and we will need to certify the
new alloy for machinability and high-radiation application.  

The $\sim100\,$W of power dissipated in these
absorbers (in Feasibility Study II at least) is within the bounds of high-power liquid-hydrogen
targets developed for, and operated in, a variety of experiments~\cite{targets}.  However, the
highly turbulent fluid dynamics involved in the heat-exchange process necessarily requires R\&D
for each new configuration. We have identified two possible approaches: a ``conventional"
flow-through design with external heat exchanger, similar to that used for high-power LH$_2$
targets, and a convection-cooled design, with internal heat exchanger built into the absorber
vessel. The convection design has desirable mechanical simplicity and minimizes the total hydrogen
volume in the cooling channel (a significant safety concern), but is expected to be limited to
lower power dissipation than the flow-through design.  To study and optimize the fluid mixing and
heat transfer properties of these absorber designs, we have been exploring ways to visualize the
flow patterns and temperature distributions within the fluid~\cite{Norem-Schlieren} and test the 
predictions of numerical flow simulations~\cite{Obabko}. To keep window thicknesses to a minimum, both designs operate just above 1\,atm of pressure. 

Various scenarios have been discussed involving substantially higher absorber power dissipation:
1)~a Neutrino Factory with a more ambitious Proton Driver (4\,MW proton-beam power on the
pion-production target instead of the 1\,MW assumed in Study-II) is a relatively straightforward
and cost-effective design upgrade~\cite{Alsharoa}, 2)~the ``bunched-beam phase rotation" scenario
of Neuffer~\cite{Neuffer-bunch} captures $\mu^+$ and $\mu^-$ simultaneously, doubling the absorber
power dissipation, and 3)~a ring cooler~\cite{Palmer-ring} would entail multiple traversals of
each absorber by each muon, potentially increasing absorber power dissipation by an order of
magnitude. If all three of these design upgrades are implemented, power dissipations of tens of
kilowatts per absorber will result. The large heat capacity of hydrogen means that such levels of
instantaneous power dissipation are in principle supportable, but much higher average heat
transfer would be needed, possibly requiring higher operating pressure and thicker windows. More work is
needed to assess the muon-cooling performance implications. 

\subsection{Other absorber materials}

Other candidate absorber materials include helium, lithium, lithium hydride, methane, and
beryllium. All other things being equal, in principle these would all give worse cooling
performance than hydrogen. For fixed $\beta_\perp$, a possible figure of merit is $(L_R\,\langle
{d}E/{d}s\rangle_{\rm min})^2$ (proportional to the four-dimensional transverse-cooling rate),
normalized to that of liquid hydrogen.  Table~\ref{tab:matl} shows that hydrogen is best by a
factor $\approx2$, although its advantage could be vitiated if thick windows are necessary.
Furthermore, for sufficiently high focusing-current density, lithium lenses could provide
substantially lower $\beta_\perp$ than is practical with
solenoids~\cite{Neuffer83,Neuffer-COOL03}, perhaps sufficient to overcome lithium's
disadvantageous merit factor. Liquids provide high power-handling capability, since the warmed
liquid can be moved to a heat exchanger. On the other hand, the higher densities of solids allow
the absorber to be located more precisely at the low-beta point of the lattice. Lithium hydride
may be usable with no windows, but means would have to be devised to prevent combustion due to possible
contact with moisture, as well as to avoid melting at high power levels. More work will be
required to assess these issues in detail.

It has been pointed out~\cite{Kaplan-NuFACT01,MCNote195} that gaseous hydrogen (GH$_2$) at high
pressure could provide the energy absorption needed for ionization cooling, with significantly
different technical challenges than those of a liquid or solid absorber. Table~\ref{tab:matl}
shows that GH$_2$ is actually a slightly better ionization-cooling medium than LH$_2$. In
addition, if the hydrogen is allowed to fill the rf cavities, the number of windows in the cooling
channel can be substantially reduced, and the length of the channel significantly shortened.
Moreover, filling the cavities with a dense gas can suppress breakdown and field emission, via the
Paschen effect~\cite{Paschen}. A small business~\cite{MuonsInc} has been formed to pursue this
idea, with funding from the U.S. Dept.\ of Energy's Small Business Technology Transfer
program~\cite{STTR}. Phase I of this program has been completed and included tests of breakdown in
gaseous helium and hydrogen at 805\,MHz, 77\,K temperature, and pressures from 1 to $\approx$50\,atm;
gradients as high as 50\,MV/m have been achieved~\cite{Johnson-PAC03}. If approved, a follow-on
Phase II will explore operation at lower frequency. Successful completion of this program could
lead to construction of a prototype gaseous-absorber cooling cell, to be tested at the MuCool Test
Area (described below) and perhaps in a future phase of the Muon Ionization Cooling Experiment
(MICE)~\cite{MICE}. Other applications for gas-filled rf cavities have also been proposed,
including rf pulse compression and six-dimensional cooling~\cite{Derbenev}.

\subsection{High-gradient normal-conducting rf cavities} 

An ionization-cooling channel requires insertion of high-gradient rf cavities into a lattice
employing strong solenoidal magnetic fields. This precludes the use of superconducting cavities.
The lattice of Fig.~\ref{fig:SFOFO} employs normal-conducting 201-MHz cavities, but R\&D is more
readily carried out with smaller, higher-frequency devices.  

Radio-frequency cavities normally contain a minimum of material in the path of the beam. However,
the penetrating character of the muon allows the use of closed-cell (``pillbox") cavities,
provided that the cell closures are constructed of thin material of long radiation length. 
Eq.~\ref{eq:cool} implies that this material will have little effect on cooling performance as
long as its thickness $L$ per cooling cell (at the $\beta_\perp$ of its location in the lattice)
has $\beta_\perp L/L_R$ small compared to that of an absorber. Closing the rf cells approximately
doubles the on-axis accelerating gradient for a given maximum surface electric field, allowing
operation with less rf power and suppressing field emission. Two alternatives have been considered
for the design of the cell closures: thin beryllium foils and grids of gas-cooled, thin-walled
aluminum tubing. As a fall-back, an open-cell cavity design was also  pursued. 

So far we have prototyped and tested a 6-cell open-cell cavity, designed at Fermilab,  and a
single-cell closed-cell cavity, designed at LBNL, both at 805\,MHz. The tests are carried out in
Fermilab's Laboratory G, where we have installed a high-power 805-MHz klystron transmitter (12-MW
peak pulsed power with pulse length of 50\,$\mu$s and repetition rate of 15\,Hz), an
x-ray-shielded cave, remote-readout test probes, safety-interlock systems, and a control room and
workshop area for setup of experiments. The cave also contains a high-vacuum pumping system and
water cooling for the cavity. To allow tests of the cooling-channel rf cavities and absorbers in a
high magnetic field or high field gradient, a superconducting 5-T solenoid with a room-temperature
bore of 44\,cm was constructed by LBNL and installed in Lab G, with two separate coils that can be
run in ``solenoid" mode (currents flowing in the same direction) or ``gradient" mode (currents in
opposite directions).  

The open-cell cavity 
was conditioned up to a surface electric field
of 54\,MV/m (on-axis accelerating gradient up to 25\,MV/m). Electron dark currents and x-ray
backgrounds were found to be large and to scale as a high power of the surface field, $E^{{\tiny \gsim}10}$~\cite{Norem}. With a 2.5-T solenoidal field applied, at 54-MV/m surface field, axially
focused dark currents ultimately burned a hole in the cavity's titanium vacuum window. This level
of background emission would preclude cavity operation in the required solenoidal field. However,
for the same accelerating gradient, the pillbox cavity operates at approximately half the surface
field, corresponding to lower background emission by a factor of order $10^{3}$. Furthermore, an
analysis of the observed emission rate in terms of the Fowler-Nordheim
formalism~\cite{Fowler-Nordheim} implies an enhancement of the emission probability by a factor of
order $10^{3}$ compared to that of a smooth, clean surface~\cite{Norem}. This suggests that an
R\&D program focused on improving the surface preparation and treatment might reap large
improvements. 

\begin{figure} 
\begin{center} 
\scalebox{0.5}{\includegraphics{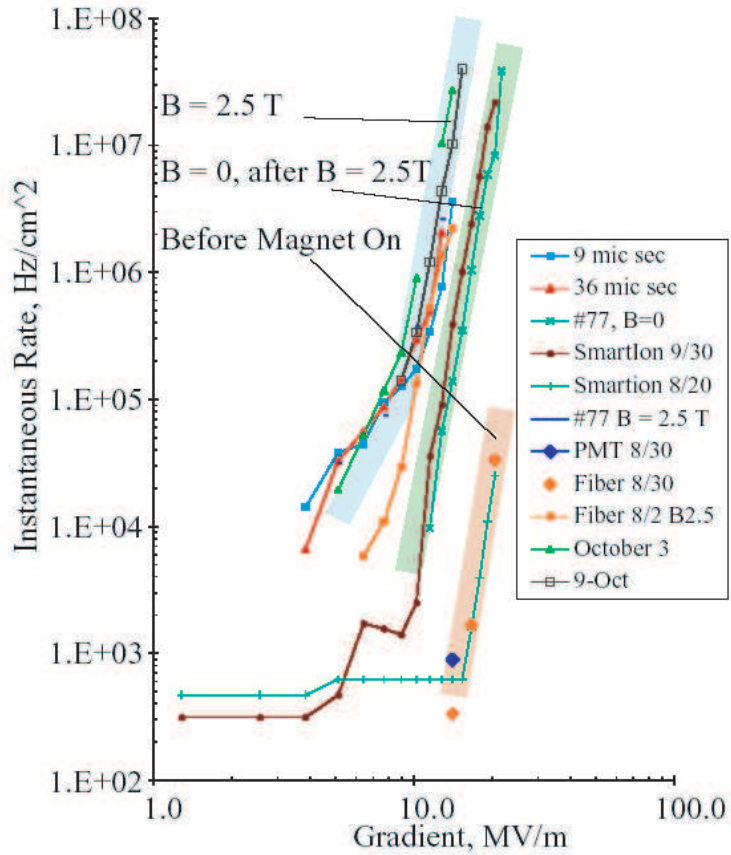}} 
\end{center} 
\caption{\label{fig:pillbox-rates}Dark-current and x-ray rates with the 805-MHz pillbox cavity 
measured under a variety of conditions with a variety of detectors as indicated.} 
\end{figure}

Tests of the closed-cell prototype are in progress~\cite{Norem-PAC03}. Initial tests with
5-mm-thick copper windows up to the design gradient of 34\,MV/m with no applied magnetic field
were carried out successfully, with an acceptable degree of multipactoring and little or no arcing. 
Upon disassembly, no damage to the
windows was observed. The thickness of the cavity's vacuum windows (1.9 cm of stainless steel)
precluded measurement of low-energy backgrounds. Thinner windows (380\,$\mu$m of Cu plus
200\,$\mu$m Ti vacuum windows) were then installed and operated with gradients above 20\,MV/m
at 2.5\,T, however occasional sparking at 2.5\,T gradually degraded performance. Some healing of
the damage was effected by reconditioning at $B=0$, but repeated cycles of damage and
reconditioning established that the highest sustainable gradient at 2.5\,T was $\approx$15\,MV/m.
Measurements of x rays and dark currents during these tests are presented in
Fig.~\ref{fig:pillbox-rates}. 

Tests were subsequently performed with TiN-coated beryllium windows. Gradients up to 18\,MV/m were
achieved at 4\,T, but again spark damage gradually degraded performance; we also encountered some
frequency instability, suggesting that the windows may have been flexing due to rf heating. When
these windows were removed for inspection, blobs of copper were found deposited on the
(otherwise undamaged) TiN
surface. An insulating TiN coating was also observed on the mounting ring, contrary to
specifications. Tests are now beginning with another set of TiN-coated Be windows having a clean
electrical connection at the mounting ring.

While our experience shows that development of closed-cell, high-gradient, normal-conducting
cavities for use in magnetic fields is not easy, there is as yet no reason to believe we will
not ultimately be successful. Exposed copper surfaces appear to be problematic at high surface
electric field. A variety of window and cavity surface preparations and coatings remain to be explored,
{\em e.g.}, TiN-coating the copper at the locations of maximum surface field. Alternatives to flat,
prestressed foils are receiving attention as well, and we expect to prototype and test several
possible solutions at 805 MHz. (Our 805-MHz pillbox-cavity prototype was designed with demountable windows
with such a test program in mind.) Design studies indicate that both precurved Be foils and grids
of gas-cooled, thin-walled Al tubes should be feasible and may be cheaper and induce less
scattering than flat foils. 

The design of a prototype 201-MHz closed-cell cavity for muon cooling is essentially complete~\cite{Li}. Since
the window R\&D is not yet completed, the cavity design accommodates a variety of cell closures.
We intend to build the first prototype in the coming year.

\subsection{Test facilities} 

To augment the Lab G facility described above, we are building a MuCool Test Area at the end of
the Fermilab Linac. This location combines availability of multi-megawatt rf power at  both  805
and 201\,MHz and  400-MeV proton beam at high intensity. Cryogenic facilities  will be provided
for liquid-hydrogen-absorber and superconducting-magnet operation. 
The underground enclosure under construction will provide the radiation shielding needed for beam
tests of absorber power handling and for high-gradient cavity testing, with the added capability
of exploring possible effects on cavity breakdown due to beam irradiation of the cavity walls in a
solenoidal magnetic field. Construction of the MuCool Test Area has progressed well and we
anticipate its utilization for absorber tests towards the end of 2003, as well as for tests of the 
prototype 201-MHz cavity when it becomes available.

The MuCool program includes engineering tests of ionization-cooling components and systems, but
not an actual experimental demonstration of ionization cooling with a muon beam. Such a cooling
demonstration (MICE) has been proposed and is discussed elsewhere in these
Proceedings~\cite{MICE}.

\section*{Acknowledgements}

I thank the organizers for the opportunity to present this work in this remarkable and beautiful setting. This work was supported in part by  the US Dept.\ of Energy, the National Science Foundation,
the Illinois Board of Higher Education, the US-Japan Agreement on High Energy Physics, and the UK Particle Physics and Astronomy Research Council.

\newpage

\begin{table}\caption{\label{tab:matl}Comparison of ionization-cooling merit factor (see text) 
for various possible absorber materials~\protect\cite{PDG}.} 
\begin{tabular}{lccc} 
\hline\hline 
 &  $\langle$d$E$/d$s\rangle_{\rm min}$ & $L_R$ & \\ 
\raisebox{1.5ex}[0pt]{Material} &  (MeV\,g$^{-1}$cm$^{2}$) & (g\,cm$^{-2}$) & \raisebox{1.5ex}[0pt]{Merit} \\ 
\hline
GH$_2$ & 4.103 & 61.28 & 1.03 \\ 
LH$_2$	& 4.034 & 61.28 &	1 \\ 
He	& 1.937 & 94.32 & 0.55 \\
LiH     & 1.94 & 86.9 & 0.47 \\ 
Li	 & 1.639 & 82.76 & 0.30 \\
CH$_4$	 & 2.417 & 46.22 & 0.20 \\
Be	 & 1.594 & 65.19 & 0.18 \\
\hline\hline 
\end{tabular} 
\end{table}

\end{document}